\documentclass[epj]{svjour}
\usepackage{graphicx}
\begin{document}
\title{Dilepton and $\phi$ meson production in elementary and nuclear
  collisions at the NICA fixed target experiment}
\author{Gy\"orgy Wolf \ and Mikl\'os Z\'et\'enyi
}                     
%
%
\institute{Institute for Particle and Nuclear Physics, Wigner Research
  Centre for Physics, Hungarian Academy of Sciences, P.O. Box 49, H-1525
  Budapest, Hungary}
\authorrunning{Gy. Wolf, M. Z\'et\'enyi}
\titlerunning{Dilepton and $\phi$ meson production...}

\date{Received: date / Revised version: date}
%
\abstract{
We argue that the NICA fixed target experiment will be able to provide
very important new experimental data on dilepton and $\phi$ meson production
in the basically undiscovered energy domain between the SIS and SPS energies.
Experimental information about elementary cross sections in this energy region
is an essential ingredient of models of nuclear collisions in the same energy
range.
} 
\maketitle
Currently, the strongly interacting matter can be accessed experimentally at 
low density 
(RHIC/Brookhaven and LHC/CERN) and at normal nuclear density (ordinary nuclear 
physics). Its properties at high densities, where the critical endpoint 
probably sits, are not known, neither experimentally nor theoretically. 
Therefore it is very important that new accelerators and detectors (NICA/MPD 
and FAIR/CBM)) are going to be built to study the properties of the dense, 
strongly interacting matter.

The chiral symmetry is expected to be restored at high densities.
The path to the restoration is not known, and may be observed by measuring 
meson properties, like masses and widths.

Dileptons are an important probe of relativistic nuclear collisions,
because they leave the hot and dense phase of the collision unaffected
by strong final state interactions and can be used to observe vector meson 
properties in dense matter via their direct decays to dileptons. 
On the other hand, the dilepton spectrum obtained 
from a nuclear collision is a complicated
superposition of many production channels coming from various stages of
the process. Therefore, drawing any reliable conclusions based on the
dilepton spectrum is difficult, and depends on understanding many
aspects of the underlying physics, like elementary cross sections, in
medium modification of particles, reaction dynamics, etc.

In recent years dilepton production was studied experimentally at RHIC, SPS, 
and at
much lower energies at GSI SIS by the HADES experiment and theoretically by 
several theoretical groups as well, a few of them to mention here:
~\cite{Wolf90,Wolf93,Lynnik,Endres,vanHees}. 
The range between the SIS and SPS energies is basically undiscovered, and the 
NICA  fixed target experiment will give a unique 
opportunity to study dilepton production in this regime. 

At low energies elementary hadron collisions were usually described by
the resonance dominance model, where in the first step of the collision
a baryon resonance is excited, which later decays -- possibly in
multiple steps -- and creates the final state particles. These kinds of
models were used to study the dilepton spectrum of nucleon-nucleon
collisions in connection with the DLS data (see
Ref.~\cite{pp_dilep_res}). However, a more or less satisfactory
agreement between theory and experiment has been reached only with the
HADES data and the new calculations in terms of effective field theory
(EFT) models \cite{OBE_pp}. A similar EFT model has been recently
applied to dilepton production in pion-nucleon collisions
\cite{Zetenyi-Wolf}.

The spectrum of baryon resonances is known only up to slightly above 2
GeV, with increasing uncertainty at high masses. The resonance model
cannot be applied to collisions above the SIS energy range not only
because the baryon resonance spectrum is unknown, but also because
multiparticle final states become more and more important. The
calculation of multiparticle production processes in EFT models is
complicated by the multi-dimensional phase-space integrals, and many
different Feynman diagrams contributing to the same final state. These
diagrams have to be added coherently, which results in a large number of
interference terms. 

\begin{figure}[htb]
  \begin{center}
    \includegraphics[width=0.5\textwidth]{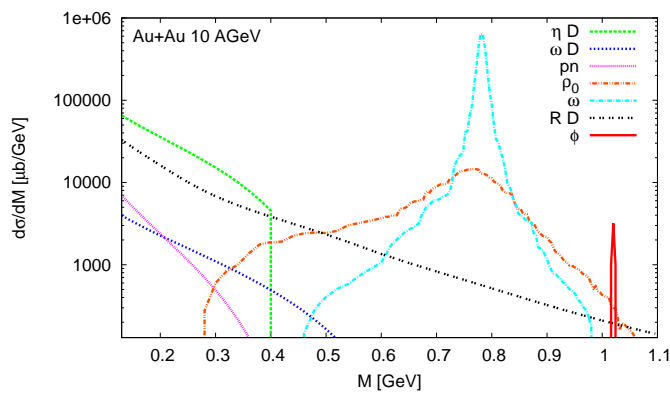}
    \caption{\label{fig:dilep}Contributions of various elementary channels
    to the dilepton invariant mass spectrum of Au+Au collisions, at 10 AGeV
    laboratory kinetic energy, as predicted by a hadronic transport model.}
  \end{center}
\end{figure}

Heavy ion collisions in similar energy ranges are usually described by
hadronic transport models. An ingredient of such models is a description
of the elementary hadronic reactions. Normally, a variant of the resonance
dominance model serves this purpose in hadronic transport codes. The reason 
of this is that EFT model calculations
contradict the philosophy of transport, where (baryon and meson)
resonances propagate as ordinary particles. This prohibits the
implementation of certain types of interference terms that appear
naturally in EFT models.

Figure \ref{fig:dilep} shows the predictions of a BUU type transport model 
(for some of the details see~\cite{Wolf90,Wolf93,Schade,WolfSpectral}) for the 
various contributions to the dilepton invariant mass spectrum of a central 
Au+Au collision in the NICA energy range of 10 AGeV. Above 0.6~GeV dilepton 
mass the 
spectrum is dominated by the direct decay of the neutral vector mesons $\rho$, 
$\omega$ and $\phi$.
For lower dilepton masses the Dalitz decays are the most important source
of dileptons. In particular, around 0.4-0.5 GeV the Dalitz decay of baryon
resonances is expected to be very important. This is because at these higher
energies a large fraction of the nucleons are excited to the resonance states.

Since the spectrum of higher baryon resonances is poor\-ly known, the 
predictions for contributions of their Dalitz decay are also rather uncertain
(see \cite{Zetenyi-Wolf2}). 
These resonances also participate in the production of 
the vector mesons, which brings in uncertainties in the vector meson channels
of dilepton production. Furthermore, the highest baryon densities 
(approximately 6 times the normal nuclear density at 10 AGeV)
are expected to be reached in heavy ion collisions in the energy range studied
at NICA.
Dilepton spectra will be influenced by the modification of hadron spectral 
functions in the dense medium. In addition, multiparticle final states
will contribution to dilepton production as well. These are
usually described in transport codes in terms of string fragmentation models. 

This shows that there must be a transition in the applied
theoretical models just above the SIS energy range, with probably a pure
EFT model (or in transport codes a resonance dominace model) for lower 
energies, supplemented by a string fragmentation
model the importance of which is increasing with energy. It is clear
that experimental input about elementary cross sections is needed in
order to test and calibrate the theoretical models at this regime of
transition. The NICA fixed target experiment is an ideal possibility to
provide these important experimental data.

Another reason why the energy range of the NICA fixed target experiment
is interesting is $\phi$ meson production. $\phi$, as a neutral vector
meson is interesting because it decays directly to the dilepton channel
and, therefore, its spectral function can be studied on the dilepton
invariant mass spectrum. Furthermore, as a particle containing hidden
strangeness, it decays dominantly to the kaon-antikaon channel. The
simultaneous study of both decay channels can contribute to a deeper
understanding of the underlying physics. In medium modification of
antikaon mass can lead to a broadening of the $\phi$. At low energy
heavy ion reactions, collision of secondary particles plays an important
role in the production of $\phi$ mesons (see \cite{phi}), therefore it's
sensitive to the reaction dynamics, the EOS etc.

In a hadronic fireball created in moderate energy heavy ion collisions,
$\phi$ mesons can be created in baryon-baryon, meson-baryon, and in 
meson-meson collisions. Of the latter, obviously kaon-antikaon collisions
are the most important.
The contributions of these channels to $\phi$
meson production in central Au+Au collisions at 10 AGeV laboratory kinetic 
energyare shown in Table~\ref{Tab:phi}. The elementary cross 
sections are obtained from one boson exchange models. Contributions of higher
baryon resonances are included, too. They actually dominate the baryon-baryon 
contributons. Their cross section is not known, can only be extrapolated from
nucleon-nucleon cross sections. Most of the $\phi$ mesons originates from 
K${}^+$K${}^-$ annihilations. Therefore any in-medium effects on the kaon and 
antikaon properties, eg. on the masses, heavily influences the dilepton spectra.

\begin{table}
\begin{center}
  \caption{
    The probability of creating a $\phi$ meson for various channels in a Au + Au 
    central collison at 10 AGeV bombarding energy (B and N denote a non-strange 
    baryon and a nucleon, respectively).}
 \label{Tab:phi} 
\begin{tabular}{ll}
\hline\noalign{\smallskip}
  channel & contribution \\
\noalign{\smallskip}\hline\noalign{\smallskip}
  BB $\rightarrow$ NN $\phi$ & $8.9 10^{-3}$ \\
  $\pi$ B  $\rightarrow$ N $\phi$ & $1.44 10^{-3}$ \\
($\rho, \omega$) B  $\rightarrow$ N $\phi$ & $6.30 10^{-4}$\\
K${}^+$K${}^-$  $\rightarrow$ $\phi$ &  $8.57 10^{-2}$\\
\noalign{\smallskip}\hline
\end{tabular}
\vspace*{0.5cm}  
\end{center}
\end{table}

The energy at SIS was not high enough to study the $\phi$ meson in
detail. In particular, it was not seen in the dilepton channel. Only a
few measurements have been performed at SIS to study $\phi$ production
via the $K^+K^-$ channel in subthreshold heavy ion collisions
\cite{FOPI}. These results by the FOPI collaboration indicate a large
$\phi$ production cross section, which is not explained by the transport
calculations of Ref.~\cite{phi}. 

At higher energies the $\phi$ meson peak in the dilepton spectrum is
expected to be more pronounced, which can be seen in the BUU 
predictions of Fig.~\ref{fig:dilep}. This means, that the NICA fixed 
target experiment would be able to study the $\phi$ meson in both decay 
channels and, therefore, could contribute to a better understanding of $\phi$ 
production around the kinematical threshold. Since the $\phi$ meson production 
is dominated by $K^+K^-$, experiments at NICA energies would enable us to study uniquely kaon 
and antikaon properties in a very dense medium (6--10 times nuclear matter 
density), which may help us understand the interior of a neutron star 
\cite{Blaschke-Weber}.

%
%

\end{document}